# Data governance: A Critical Foundation for Data Driven Decision-Making in Operations and Supply Chains

Xuejiao Li[1], Yang Cheng[1,*], Charles Møller[2]

1: Department of Materials and Production, Aalborg University, Denmark

2: Department of Mechanical and Production Engineering, Aarhus University, Denmark

* Corresponding author, cy@business.aau.dk, cy@mp.aau.dk

**Abstract:** In the context of Industry 4.0, the manufacturing sector is increasingly facing the challenge of data usability, which is becoming a widespread phenomenon and a new contemporary concern. In response, Data Governance (DG) emerges as a viable avenue to address data challenges. This study aims to call attention on DG research in the field of operations and supply chain management (OSCM). Based on literature research, we investigate research gaps in academia. Built upon three case studies, we exanimated and analyzed real life data issues in the industry. Four types of cause related to data issues were found: 1) human factors, 2) lack of written rules and regulations, 3) ineffective technological hardware and software, and 4) lack of resources. Subsequently, a three-pronged research framework was suggested. This paper highlights the urgency for research on DG in OSCM, outlines a research pathway for fellow scholars, and offers guidance to industry in the design and implementation of DG strategies.

**Keywords:** Data issues, Data governance, Manufacturing sector, Operations and supply chain

**Acknowledgement:** This study is supported by MADE FAST (MADE -- Flexible, Agile, and Sustainable production enabled by Talented employees) Project (470100), Jiangxi Double Thousand Plan, and "Strengthening the digitalization of businesses in Eastern Europe – a micro and macro-level approach" funded by the European Union – NextGenerationEU project and the Romanian Government.

# 1. Introduction

With the rise of Industry 4.0, digital transformation has become a main theme in the industrial sector. Industrial companies have been making efforts to adopt more digital technologies in their operations and supply chains (O&SC). This further leads to the continuous generation of numerous and complex data at a high speed (Reinsel et al., 2018). While such "big" data stimulates companies move towards data-driven decision making, it also becomes an obstacle for such journey. When companies use their data to support decision making in O&SC, they suffer from a range of challenges including low data quality marked by missing data, duplication, erratic labeling, lack of data description, various unlinked systems for data storage, no data access, and no standardized data application processes, among others. Unless these issues are solved, it is meaningless to adopt any business intelligence systems or data science technologies, like big data analysis (BDA), machine learning (ML), or artificial intelligence (AI), and companies cannot truly move to data-driven decision making. To tackle these complex issues, data governance (DG) emerges as a pivotal solution.

These issues indicate a need for effective data governance (DG), which refers to the exercise of authority and control over the management of data asset. It aims at implementing a corporate-wide data agenda, maximizing the value of data assets in an organization, and managing data-related risks (DAMA International, 2015). Because of its potential benefits, current studies in the information system (IS) field have paid certain attention to DG and built the research foundation. However, the studies in the IS field do not provide specific support for DG in O&SC decision-making, while existing research in the operations and supply chain management (OSCM) field lacks a comprehensive DG framework or approach.

Facing both practical necessity and theoretical vacancies, this paper aims to drawing the attention of the O&SC field to data issues, bring the topic of DG to the forefront, and establish a pathway for future research about DG in the O&SC field.

The structure of this study is organized as follows: Section 2 elucidates the current state of the art in DG research. In Section 3, the methodology employed in this study is outlined. Section 4 proposes a research pathway and framework for future investigations. Finally, Section 5 encapsulates the conclusions drawn from this study along with its limitations.

## 2. Literature review

*From data management to data governance*

DG is often used interchangeably with data management (DM), but they are essentially different in terms of driving force and purpose. DM is the daily operations and behaviors related to the technical management of data, aiming to ensure the availability of data to meet the requirements of all applications but caring less about how the data is used. Today, it is essentially the responsibility of IT department. Compared to the data centric characteristic of DM, DG is more business centric, being the process of controlling data to make it more usable and compliant with legislation and best practices. It aims to summarize, integrate, and manage data of different structures, types, and sources, so business people can easily identify and find the data resource they need for decision-making. Thus, DG requires close cooperation between IT and business departments and emphasizes the construction of outlines, policies, and procedures that guide the use, exchange and sharing of data in the enterprise by clarifying who can take what actions with what data, and when, under what circumstances, using what methods (Plotkin, 2020).

*Research on DG in the field of IS and from industry*

The field of IS has a long history of studying how to manage data, with the focus shifting from DM to DG. Early studies were more technical, focusing on specific aspects, e.g., metadata and data quality management, data quality, and data lifecycle (e.g., Donaldson and Walker, 2004). However, it has been gradually recognized in practice that traditional technology- or IT-driven DM projects are difficult to be implemented and not helpful for solving data-related problems in business. Thus, more scholars took the other way around, i.e., focusing on top-level design of DG from the perspective of strategy and organization for realizing DM goals. Hence, more research has been done to propose generic and theoretical DG frameworks (e.g., Alhassan et al., 2016; Abraham et al., 2019), which specify cross-functional frameworks for managing data as a strategic asset and decision rights and accountabilities for an organization's decision-making about its data, and formalize data policies, standards, and procedures. These frameworks are similar to those developed by practitioners, e.g., IBM, Gartner, and Data Governance Institute. However, all the frameworks are theoretical and less operationalizable; they are useful to pinpoint key aspects to be considered, but do not talk about tangible actions to create real change; and they are about data in general, but do not govern data for any specific decision and cannot be applied directly in practice to guide companies' behaviors. Last but not least, they also fail to address the data generated in O&SC environment, even this data is more complex as generated by different machines, with different formats, from different stakeholders. Hence, the discussions reflect the needs of IT function more for better managing data in its life cycle from acquisition to elimination, rather than helping companies use data to make better decisions in O&SC.

*Research on DG in the field of OSCM*

Meanwhile, research on DG in the field of O&SC has not received sufficient attention, with only a handful studies identified. These studies do not directly explored DG as a whole, or provide comprehensive, guiding DG framework for O&SC, but ends to emphasize specific practical problems within the operational processes of companies. These studies touch upon various data-related issues when developing solutions such as big data analytics (BDA), machine learning (ML), or AI-based approaches to support decision-making.

The majority of attention was directed towards factory-level production systems, particularly about new product development and maintenance. For new product development, the existing studies have discussed e.g., data retrieval, integration, sharing, and quality for supporting product development (e.g., Do, 2015). For maintenance, the relevant discussions involved e.g., data integration to improve maintenance management, data quality evaluation and improvement for prognostic modeling and fault detection (e.g., Omri et al., 2021). In contrast, less research moved beyond factory walls, with a few examining the effects of analytical capabilities, data quality, and real-time data capabilities on supply chain performance (e.g., Jalil et al., 2011; Shehab et al., 2013; Hazen et al., 2014; Oliveira and Handfield, 2019) and exploring the electronic regulation of data sharing and processing for supply chain security (Epiphaniou et al., 2020). Last, some studies have also addressed e.g., data quality, data architecture and storage, and the analysis of machine-generated data in a general O&SC context without linking with specific decisions (e.g., Villalobos et al., 2020). In short, these existing studies imply that data, analytics, and decision making are interlinked, but they mainly focus on solving individual data-related problems (e.g., data quality, data integration), rather than developing a holistic framework or approach to guide companies for better DG. Hence, the discussion on DG in the O&SC field is still scant, sparse, fragmented, and less systematical.

*Literature synergy and research questions*

DG has been relatively well addressed in the field of IS. The existing studies focus more on developing generic and theoretical frameworks. However, the disadvantage of the frameworks is that they are theoretical and less operationalizable, lacking tangible actions for creating real change, might not be applied directly to facilitate O&SC functions. Meanwhile, DG has been less discussed in the field of OSCM, and the existing research is scant, sparse, and fragmented. These studies do not specifically focus on DG, but refer to DG-related concepts, e.g., data quality, data integration, when elaborating their BDA, ML or AI-based solutions for production problems. In these studies, there is a pattern that data, analytics, and decision making are linked, which can be adopted to better apply DG to practical operations. However, DG research in the OSCM domain mainly focus on solving individual problems but lacks a comprehensive framework and high-level guidance for companies. Therefore, this study aims to address these gaps by investigating the question of "How to design a DG framework in O&SC to leverage data for decision-making".

## 3. Case study
### 3.1 Methodology

In order to answer the research question, the prerequisite is to understand the business and data usage in real O&SC context. Thus, three case studies are conducted in this section. The methodology is as follows.

*Case Selection:*

In order to make the case more rigorous and representative, we chose three cases covering different scopes within a supply chain: within-plant, supply (upstream of supply chain), and distribution (downstream of supply chain). The case companies are all large multi-national manufacturing companies. Large companies may encounter more diverse and complex data issues due to their scale and the corresponding amount of data, which will provide the research with richer materials.

*Data Collection:*

The data was collected between October 2021 and March 2023 through semi-structured interview and documentation. Interviews were conducted with managers and employees who were involved in the data utilization projects including the senior and middle managers, IT consultants, and operators. Since this study focused on the application of data for decision-making, questions in relation to the business problems, the decisions to be made, how the data is used in the business, and what data issues are facing by the business were asked during the interviews. Interviewees were clearly informed that they had the right to withdraw at any time during the interviews. The interviews were audio-taped with the consent from interviewees and the confidentiality of their response was confirmed.

*Data Analysis:*

When analyzing the data, we initially organized the business situation of the case companies and created workflow diagrams. Subsequently, we examined the data usage in each workflow step, delineated the data flow, and documented any encountered data issues. Finally, we proceeded to analyze the data issues, and the 5 WHY analysis technique is employed to identify the causes of the data issues.

**3.2 Case introduction**

Case 1 focused on the bottling plant of a beverage company in Denmark. The plant has invested significantly to digitalize its production lines. Large amounts of data have been collected from sensors and machines, which is believed important for e.g., measuring and improving production line efficiency, supporting product quality control, and facilitating machine maintenance.

Case 2 explored the upstream of a multinational wind energy company in Denmark, in collaboration with its supply chain department. The department is responsible for strategic and tactical planning of the group's supply chain. Large amount of data from every aspect in the supply chain are expected to be used in the planning of demand, production, transportation and component supply.

Case 3 delved into the downstream of a household goods company, i.e., its distribution center located in Denmark. The center's responsibility is to pack and distribute goods to downstream customers, where data is believed to play a vital role in daily packing, distribution, and performance analysis.

**3.3 Case analysis**

*3.3.1 Examination of real-life data issue*

Three cases offered insights into practical data usage. We identified challenges in various aspects of data collection, processing, and utilization. To facilitate a structured examination of these data problems, we utilized the knowledge areas of a well-known framework proposed in The DAMA Guide to the Data Management Body of Knowledge (DAMA, 2015). The choice of this particular framework was based on its comprehensiveness in comparison to other DG frameworks. Table 1 provides a detailed list of data-related problems and their potential effects.

Table 1. List of data issues

| DM knowledge areas | Data issue | Potential effects |
|---|---|---|
| **Data quality** | • 1. Missing data (reported by case 1)<br>• 2. Duplication (reported by case 1)<br>• 3. Wrong value (reported by case 1)<br>• 4. Messy labeling (e.g., different data with the same name) (reported by case 1)<br>• 5. Name convention does not make sense or no data descriptions/ name convention (reported by case 1)<br>• 6. Data sheets that have no relevance for the company (reported by case 1) | • Inaccurate insights<br>• Flawed decision-making<br>• Unreliable analysis<br>• Biased results<br>• Compromised validity<br>• Incomplete information<br>• Unreliable predictions<br>• Hindered performance evaluation |
| **Data integration & interoperability** | • 7. Multiple data sources increase the time of searching and retrieving data (reported by case 1, 2 and 3)<br>• 8. Difficulties on obtaining self-extracted raw data or accessing to stored but hidden data (reported by case 1)<br>• 9. Data silos in systems (reported by case 1)<br>• 10. Shadow IT (reported by case 2 and 3)<br>• 11. No requirement of explaining what the purpose of the request for a specific data (reported by case 2)<br>• 12. "One-end" data flow (reported by case 2) | • Data inconsistency<br>• Fragmented insights<br>• Inefficient processes<br>• Limited collaboration<br>• Reduced accuracy<br>• Hindered data-driven decision-making |
| **Data storage & operation** | • 13. Storage data only on personal computer (reported by case 2 and 3)<br>• 14. Data is not archived or deleted after using (reported by case 2)<br>• 15. Nonstandard data sources make the data less reliable, e.g., some data comes from humans (reported by case 2)<br>• 16. Difficulties on using excessive amount of data (reported by case 1, 2 and 3)<br>• 17. Time consuming manually compiling work (reported by case 1, 2 and 3)<br>• 18. Information systems setting is old and lack of maintenance (reported by case 1)<br>• 19. Systems are slow (reported by case 1) | • Data corruption<br>• Limited accessibility<br>• Compromised data security<br>• Increased storage costs<br>• Inefficient data processing |
| **Data warehouse & business intelligence** | • 20. Data analytics rely on one person's own knowledge and experience (reported by case 2 and 3)<br>• 21. Inflexible data analytics (e.g., in case 1 and case 2, inflexible reports creation from data were reported)<br>• 22. No standardized decision-making principle based on data (e.g., in case 3, it is reported that there were no standardized answers for determining an operator's performance) | • Inaccurate reporting<br>• Unreliable insights<br>• Delayed decision-making |

*3.3.2 Data issue analysis and cause identification*

Each case highlighted data utilization issues in decision-making. To delve deeper, we conducted a root cause analysis focusing on the 22 data problems in Table 1 using the 5 WHY technique. This technique involves asking "why" five times to uncover the problem's nature and solution (Taiichi Ohno, 1988). The results, as shown in the Appendix, were reviewed and confirmed by practitioners from the case companies. The results are showing in Table 2.

Table 2. 5 WHY analysis of data issues

| Data issue | Answer of why1 | Answer of why2 | Answer of why3 | Answer of why4 | Answer of why5 |
|---|---|---|---|---|---|
| **Missing data (Data entry)** | Some data has not been entered into the system | The person responsible for entering the data forgot or was not aware of the requirement to enter the data | Poor communication or training, or the data entry requirement was not clearly defined or documented | Lack of standardization in data entry processes | Lack of awareness, resources, or investment in data governance |
| **Data duplication (Data storage, Data retrieval)** | The same data was entered into the system multiple times | Lack of communication and coordination among different departments or individuals responsible for managing the data | No centralized database or system for storing and accessing the data | No standardization of data storage and retrieval, no clear ownership or accountability for the data | Lack of awareness, resources, or investment in data governance |
| **Wrong data value (Data entry)** | Incorrect value was entered into the system | Human error or lack of understanding of the correct value | Lack of training, knowledge, or experience with the data or system, or high workload, distractions, or other external factors | Insufficient workforce, or poor oversight of data entry processes | Lack of investment in data governance resources or processes |
| **Messy data labeling (Data labeling)** | The labels used to describe the data are inconsistent, incorrect, or incomplete | The people responsible for labeling are not aligned or not aware of how to label data | Poor communication or training, or the data labeling process is not well-defined or documented | Lack of standardization and oversight in data labeling processes | Lack of awareness, resources, or investment in data governance |
| **Wrong or no data name convention (Data naming)** | Inconsistent, unclear, incorrect, or no data names were entered | The person responsible for naming the data forgot or was not aware of the requirement to name the data | Poor communication or training, or the data naming requirement was not clearly defined or documented | Lack of standardization in data naming processes | Lack of awareness, resources, or investment in data governance |
| **Data sheets are not relevant to the company (Data sheets management, Data maintenance)** | The data sheets are either outdated, inaccurate, not aligned with the company's current goals, or not reflective of the company's current processes | Lack of regular maintenance and review of data sheets | Lack of clear guidelines, training, or communication on data maintenance | Lack of standardization in data maintenance | Lack of awareness, resources, or investment in data governance |
| **Multiple data sources (Data sources)** | Different departments use different systems or databases | There has been no consolidation of systems | Lack of resources, investment, or awareness | Lack of centralized data governance | Lack of awareness, data-driven culture, leadership, and advocacy for |

| | | | of the benefit of the consolidation of systems | | centralized data governance initiatives |
|---|---|---|---|---|---|
| **Difficulties on obtaining wanted data (Data storage, Data sharing)** | The data is not accessible, or the data is not in the desired format or structure | Data storage or data sharing is not consistent for the desired purpose | Lack of clear data storage and sharing requirements or guidelines | Lack of standardization in data storage and sharing | Lack of awareness, resources, or investment in data governance |
| **Data silos in systems (Data storage, Data sharing)** | Data stored in separate, isolated systems | Technical limitations | Systems or platforms are not compatible with each other | Vendor lock-in, legacy systems, or a lack of universally accepted standards or protocols for interoperability | Old systems design without considering future compatibility, complex and costly to develop and maintain interoperability protocols |
| **Shadow IT (Data storage, Data sharing)** | Employees use their own tools or applications | To be more productive or efficient | The company's IT systems are too slow, outdated, or restrictive | Not enough resources, budget, or expertise to keep up with the latest technology trends and innovations | Lack of awareness or budget to prioritize IT investments |
| **No explain of requesting specific data (Data sharing)** | Employees may not understand the importance of providing context for the data request | No corresponding policy or guideline | No clear ownership or accountability for data | The company may not understand the value of data ownership and accountability | Lack of leadership or vision around data governance |
| **One end data flow (Data sharing)** | No feedback mechanisms for data | No corresponding policy or guideline | No clear ownership or accountability for data | The company may not understand the value of data ownership and accountability | Lack of leadership or vision around data governance |
| **Storage data only on personal computer (Data storage, Data sharing)** | Convenient and readily accessible | Full control over data and access it at any time | Lack of clear data storage and sharing requirements or guidelines | Lack of standardization in data storage and sharing | Lack of awareness, resources, or investment in data governance |
| **Data is not archived or deleted after using (Data storage, Data lifecycle)** | The value of archiving or deleting data after using is not seen | No clear data retention policy or data lifecycle management guidelines, or lack the knowledge or expertise to implement such policies and guidelines | No dedicated resources responsible for managing data lifecycle | May not prioritize data lifecycle management | May not have a clear understanding of the potential risks and consequences of not managing data properly |
| **Nonstandard data sources (Data sources)** | Data sources may not be properly verified, and the data may not be standardized or normalized | No clear understanding of what constitutes reliable data sources or standardization practices | Not have the necessary resources, expertise, or technology to properly verify data sources and standardize data | Not prioritize standardizing data collection and integration, or lack of support from key stakeholders | Not understand the importance, or may not have the necessary training or education |
| **Difficulties on using excessive amount of data (Data collection)** | Too much data to process and analyze | The data may be unstructured, inconsistent, or contain irrelevant information | Data is gathered from multiple sources, is not properly labeled, or lacks clear definitions or categories | Lack of standard process for data collection, labeling or classification | No clear ownership or accountability for data |
| **Time consuming manually compiling work (Data storage, Data processing)** | Manual compilation and processing of data is required | The data is stored in multiple locations, different formats, or is not standardized | Data is collected and managed by different teams or individuals who may have different processes and tools for data processing and analyzing | No standardization or centralized data governance | No clear ownership or accountability for data, lack of resources or technology to support data governance, or limited collaboration between different teams |
| **IT systems are old and lack of maintenance (IT systems)** | IT systems have not been upgraded for a long time | Lack of resources or budget to keep the systems up to date | Not prioritize IT investments | Lack of awareness or IT investments are not aligned with business goals | Lack of clear communication channels, collaboration or alignment on priorities, or a lack of a strategic roadmap for IT |
| **Information systems are slow (IT systems, Infrastructure)** | Information systems take too long to process requests | Too much traffic on the network | The infrastructure is not designed to handle the current volume of requests | The organization may not have anticipated the growth in demand for the system, or has not invested in upgrading the infrastructure | Lack of proactive planning and communication, or lack of awareness or prioritization of the importance of investing in IT infrastructure |
| **Data analytics rely on one person's knowledge and experience (Data analytics)** | That person may have a unique skill set or expertise | Other team members do not feel confident to effectively analyze and interpret the data | Other team members may not have received sufficient training or have access to the necessary tools and resources to support data analysis | Lack of investment in professional development | Lack of understanding of the potential value and importance of data analytics in decision-making |
| **Inflexible data analytics (Data analytics)** | Current data analytics tools and systems have limited functionalities and do not allow for customization | Current data analytics tools and systems were designed to serve a general purpose and cannot be easily | It was assumed that a general-purpose tool would be more cost-effective and easier to maintain | Customizing tools to fit specific needs requires a lot of updates and maintenance | The cost and effort of updates and maintenance may be too high |

| | | customized to fit specific needs | | | |
|---|---|---|---|---|---|
| **No standardized decision-making principle based on data (Data collection, Data management, Data usage)** | No clear agreement on what and how data should be used to inform decisions | Different priorities and perspectives on which data is most important, or have no idea of how to use data | No centralized data governance framework that sets standards and guidelines for data collection, management, and usage | Lack of awareness, resources, or investment in data governance | Lack of data-driven culture, leadership and advocacy for data governance initiatives, or competing priorities and budget constraints. |

Subsequently, we conducted keyword extraction and analysis on the 110 underlying reasons behind these 22 data issues. Initially, we removed common words that carry little meaning, such as articles, prepositions, and conjunctions, in order to focus on significant information. Next, we identified and tagged as keywords the terms directly related to specific measures, as these words are crucial for addressing the issues. For example, in the sentence "*...poor communication or training, or the data entry requirement was not clearly defined or documented...*" we tagged "communication," "training," and "data entry requirement" as keywords. Finally, we performed topic modeling on these keywords to discover latent topics within them.

Through this analysis, we identified four main topics that emerged from the keywords. These topics are related to 1) human factors, 2) written rules and regulations, 3) technological hardware and software, and 4) resources. Among them, human factors refer to the qualities of employees and managers themselves, as well as the interaction between individuals. These include awareness, understanding, knowledge, experience, expertise, leadership, advocacy, the culture fostered within the organization, the training and education received within the organization, and communication and collaboration among individuals. The core of improving these factors lies in human. Written rules and regulations refer to well-formulated and documented textual files, such as requirements for data entry, and regulations for the standardization of data sources. The core of improving this aspect lies in the necessity of having documented records. Technological hardware and software refer to the infrastructure, systems, platforms, tools, applications, and other necessary components required to leverage technological support. Lastly, resource-related factors encompass materials, workforce, financial resources, and other resources.

Table 3 demonstrates the specific keywords included under each topic. These findings are instructive for us to design the research pathway of DG in the O&SC. The pathway designed in the next chapter will focus on these four topics.

Table 3. Keywords and types of causes of problems

| Types of causes | Keywords | Examples |
|---|---|---|
| **1. Human factors** | Leadership, advocacy, data-driven culture, training, education, awareness, understanding, knowledge, experience, expertise, communication, coordination | <ul><li>Poor communication or training</li><li>Human error or lack of understanding of the correct value</li><li>The people responsible for labeling are not aligned or not aware of how to label data</li><li>Lack of awareness, data-driven culture, leadership, and advocacy for centralized data governance initiatives</li></ul> |

| 2. Written rules and regulations | Requirements, guidelines, documented process, policy, standardization, oversight, ownership, accountability, protocols, feedback mechanisms, strategic roadmap, framework | <ul><li>Lack of standardization in data entry processes</li><li>Data naming requirement was not clearly defined or documented</li><li>Lack of universally accepted standards or protocols for interoperability</li><li>No clear ownership or accountability for data</li></ul> |
|---|---|---|
| 3. Technological hardware and software | Databases, systems, platforms, tools, applications, infrastructure | <ul><li>No centralized database or system for storing and accessing the data</li><li>Systems or platforms are not compatible with each other</li><li>The company's IT systems are too slow, outdated, or restrictive</li><li>The infrastructure is not designed to handle the current volume of requests</li></ul> |
| 4. Resources | Workforce, investment, competing priorities, budgets | <ul><li>Lack of investment in data governance resources or processes</li><li>Not enough resources, budget, or expertise to keep up with the latest technology trends and innovations</li></ul> |

## 4. DG in O&SC pathway

Through literature research and case study, this study has synthesized existing concepts, methods, and frameworks of DG, while also identifying and analyzing real-world data issues in the O&SC context. This section aims to developing a robust DG pathway for OSCM research which can successfully integrate and address the four critical topics mentioned in section 3.3. To achieve this goal, we propose a three-pronged approach: the Top-down approach, the Bottom-up approach, and technical support. The following sections provide a comprehensive explanation of each approach.

### 4.1 Top-down approach

A top-down approach starting from the management level is suggested. This approach addresses the causes of 1) *leadership, advocacy, data-driven culture, training,* and *education* of the *human factors*, 2) *written rules and regulations*, and 4) *resources*.

This approach emphasizes the role of management in driving effective DG. It involves strong *leadership* that promotes the importance of DG, advocates for its implementation, fosters a *data-driven culture* within the organization, and provides adequate *training and education*. To do so, senior management should promote and prioritize the use of data in decision-making. This can be achieved by, for instance, providing incentives for employees to use data in their daily work; establishing performance metrics that measure the effectiveness of DG initiatives, etc.

This approach also entails establishing clear written rules and regulations that outline the DG *strategy, policies, principles*, and *processes* with the lead of senior management. A DG committee with senior leaders is recommended, which brings together the business and technical sides. The DG strategy should align with the overall business strategy, and the policies, principles, and processes should cover various aspects of DG, according to actual circumstances.

Last, this approach requires proper resource allocation. Senior management must prioritize DG's strategic importance and address resource gaps. In terms of the technical side, it is imperative to allocate resources towards building and maintaining a robust DG infrastructure. In terms of expertise, recruiting skilled DG professionals and investing in training and development are important.

**4.2 Bottom-up approach**

A bottom-up approach prioritizes business and operational levels, addressing *awareness, understanding, knowledge, experience, expertise, communication, coordination* of the *human factors*. It promotes individual and team engagement in DG, emphasizing raising awareness among employees about DG's importance, enhancing their DG knowledge and expertise, and improving communication and coordination, particularly between business and technical personnel.

The bottom-up approach starts from the business problems and helps utilizing data in the business. The core idea of this approach is that DG should be business centric. It challenges the traditional logic of using data for decision-making: obtaining data => analyzing data => making decisions. In simple terms, this logic can be expressed as "*what data is there determines what decisions can be made based on the data*". This is a reactive rather than proactive logic, which can lead to problems such as aimless data collection and redundancy. Thus, essentially focusing on DG, future research could adopt a reverse engineering approach, i.e., starting from O&SC decisions, determining what data analysis technologies are needed for making different decisions and how, and then specifying requirements on data as the input.

Following the above logic, business personnel should take the lead while technical personnel playing a supporting role. It commences with business personnel comprehending the O&SC problems and decisions to be made, followed by data analysis facilitated by technical personnel. Through effective communication and collaboration, this approach promotes a bottom-up exchange of ideas, experiences, and feedback, which can in turn improve the management level DG design in the Top-down approach.

**4.3 Technical support**

Finally, it is crucial to adopt a dedicated technical support approach that specifically focuses on *technological hardware and software*. This approach aims to establish *robust infrastructures, consolidated systems, compliant platforms, and effective tools*, all of which are essential for facilitating DG activities across the organization. Enterprises should consider modifying and redesigning the technical components according to their business needs.

For example, according to our investigation, data acquisition and integration between different systems can be a significant issue. Based on some best practices experience, it is recommended to establish a platform that encapsulates data and other digital resources by business domains, which can be called as Data Middle Platform (Alibaba, 2017). The data middle platform is a structure that lies between the data backend and the business front-end. On the platform, data and the corresponding data analysis tools are divided and packaged according to business domains. The advantage of this approach is its Packaged Business Capability (Gartner, 2019). Business personnel can quickly and accurately find the required data as data and tools are organized by business domains. The knowledge and experience generated by using data can also be accumulated on this platform. As business changes, resources on the platform will also be continuously updated. Therefore, by establishing this platform, it not only improves business processing efficiency, but also avoids the problem of knowledge being lost when employees are transferred. Unlike traditional data warehouses, where data is generated based on processes, the concept of the data middle platform focuses more on data as the core and emphasizes the relationship between data and business, while providing opportunities to maximize the value of data.

The data middle platform is just an example derived from industry practices, and further research in this area could focus more on the practical problems, technological developments, and idea generation in industry, and the integration with the academic community.

**4.4 A framework and research agenda**

The three approaches above should ultimately create a self-contained cycle. The management level leads and influences the business, operational, and technical levels, while the latter offers feedback to the management regarding problems encountered in practical operations. The depiction of these three approaches and their interrelationships can be illustrated in Figure 1.

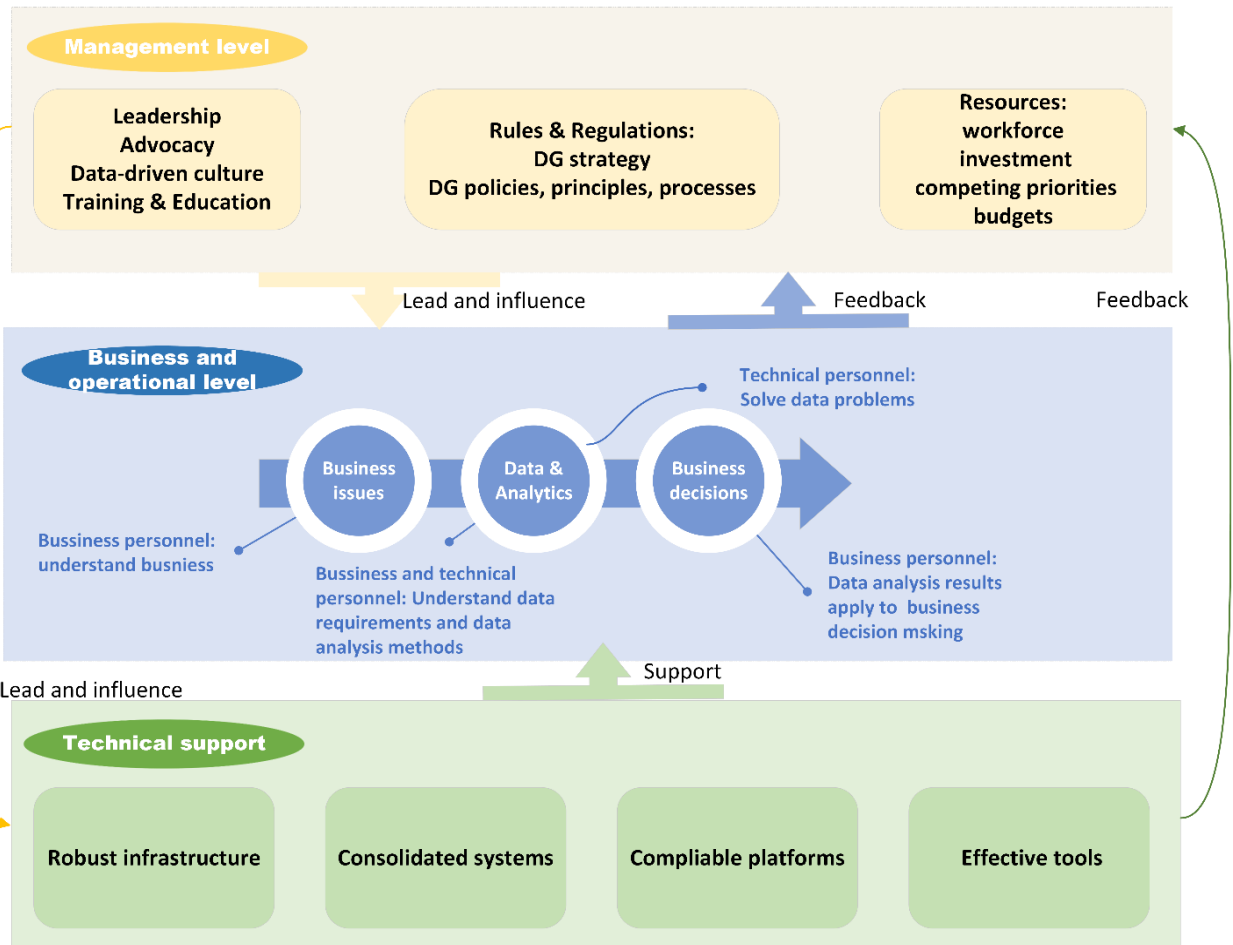

Figure 1. A framework of the three suggested pathways

Figure 1 can also be used to guide companies for initiating the establishment of their DG systems, but there do exist questions in each of the approaches that require further investigation. Thus, we present a research agenda in Table 4, summarizing the potential research questions and calling the attention from academia.

Table 4. Research agenda on DG research in the O&SC

| Approach | Research questions |
|---|---|
| **Top-down** | 1. What is the optimal leadership role for senior executives in the context of data governance?<br>2. What organizational adaptations are necessary to cultivate a data-driven culture effectively?<br>3. How can organizations formulate data strategies that harmonize seamlessly with their overarching business strategies?<br>4. Within the realm of data governance, which specific domains warrant investment, and what methodologies should be employed to assess the efficacy of these investments?<br>5. What requisite knowledge and experiential proficiencies are essential for unlocking the full potential of DG, and how can these competencies be acquired and honed?<br>6. What are the fundamental principles and processes underpinning the development of data-related policies? |
| **Bottom-up** | 1. What formal communication and collaboration mechanisms are imperative for establishing synergy between business and technical personnel? |

| | 2. How can organizations systematically determine the diverse data requisites essential for rendering informed decisions within the domain of Operations and Supply Chain?<br>3. What methodologies and processes can be employed to make data useable for advanced algorithms to assist enterprises in decision-making?<br>4. What is the optimal strategy for the formulation and implementation of data-related policies?<br>5. How can organizations institute a robust support and feedback mechanism that bridges employees with top-level management, thereby efficiently addressing data-related challenges? |
|---|---|
| **Technical support** | 1. What strategies and methodologies should manufacture enterprises employ to optimize their infrastructure and effectuate the consolidation of systems, particularly tailored for the purpose of conducting robust big data analytics?<br>2. How can organizations systematically conceptualize, design, and implement universally applicable data analysis tools to empower business personnel with data analysis autonomy?<br>3. What are the critical elements of an infrastructure designed for conducting rigorous big data analytics that is custom-tailored to the unique requisites of manufacturing enterprises?<br>4. By what means can organizations create mechanisms that foster seamless consolidation and communication among diverse systems operating within manufacturing contexts?<br>5. How can academic research endeavors focused on data governance be efficiently translated into pragmatic and industry-operable methodologies and frameworks? |

## 5. Conclusion and limitation

This study is inspired by the prevailing issue that numerous companies are commonly experiencing. Although they make significant investments and efforts in digital and intelligent transformation, the results are not desirable due to data issues. To address this issue, our study brings the topic of DG to the forefront of the scholars in the O&SC field. We exanimated and analyzed real-life data issues in O&SC and identified four causes underlying these data issues: 1) human factors, 2) lack of written rules and regulations, 3) ineffective technological hardware and software, and 4) lack of resources. In response to these identified causes, we propose a DG research pathway designed to effectively mitigate each of these causes. This pathway adopts a three-pronged approach, accompanied by the delineation of potential research concerns associated with each approach. The pathway can also extend its utility beyond academia, offering guidance to industry in the design and implementation of DG strategies.

In general, the exploration and enrichment of DG in the OSCM domain requires massive and long-term work, as well as continuous improvements and additions as the industry develops. Hence, it cannot be done by a single work, but relies on combined effort from multiple scholars in the OSCM domain. This study only sheds light on the research pathway and points out future directions on DG topic within organizations. Future research can further investigate the issues mentioned above as well as exploring effective methods to study and implement DG cross-organizationally. Furthermore, along with DG methods and theories maturing, the focus should be gradually shift to incremental data management and the mining potential value of data in the future, rather than dealing with data issues as current situation.

**Appendix. Root cause analysis on the 22 data issues using the 5 WHY technique.**

1. Missing data

W1. Why is there data missing?

   A1. Because some data has not been entered into the system.

W2. Why hasn't the data been entered into the system?

   A2. Because the person responsible for entering the data forgot or was not aware of the requirement to enter the data.

W3. Why did the responsible person forget or not know about the requirement to enter the data?

   A3. Because of poor communication or training, or because the requirement was not clearly defined or documented.

W4. Why was there poor communication or training, or why was the requirement not clearly defined or documented?

   A4. Because of a lack of standardization in data entry processes.

W5. Why is there a lack of standardization in data entry processes?

   A5. Because of a lack of awareness, resources, or investment in data governance?

2. Data duplication

W1. Why is there data duplication?

   A1. Because the same data is entered into the system multiple times.

W2. Why is the same data entered into the system multiple times?

   A2. Because there is a lack of communication and coordination among different departments or individuals responsible for managing the data.

W3. Why is there a lack of communication and coordination among different departments or individuals responsible for managing the data?

   A3. Because there is no centralized database or system for storing and accessing the data.

W4. Why is there no centralized database or system for storing and accessing the data?

   A4. Because there is no standardization of data storage and retrieval, and no clear ownership or accountability for the data.

W5. Why is there no standardization of data storage and retrieval, and no clear ownership or accountability for the data?

   A5. Because there is a lack of awareness, resources, or investment in data governance.

3. Wrong data value

W1. Why is there a wrong data value?

   A1. Because the incorrect value was entered into the system.

W2. Why was the incorrect value entered into the system?

   A2. Because of human error or a lack of understanding of the correct value.

W3. Why did the human error or lack of understanding occur?

   A3. Because of a lack of training, knowledge, or experience with the data or system, or because of a high workload, distractions, or other external factors.

W4. Why is there a lack of training, knowledge, or experience with the data or system, or high workload, distractions, or other external factors?

   A4. Because of insufficient workforce, or poor management and oversight of data entry processes.

W5. Why is there insufficient workforce, or poor management and oversight of data entry processes?

   A5. Because of a lack of investment in data governance resources or processes,

4. Messy data labeling

W1. Why is the data labeling messy?

   A1. Because the labels used to describe the data are inconsistent, incorrect, or incomplete.

W2. Why are the labels inconsistent, incorrect, or incomplete?

   A2. Because the people responsible for labeling are not aligned or not aware of how to label data.

W3. Why are the people responsible for labeling are not aligned or not aware of how to label data?

   A3. Because of poor communication or training, or the data labeling process is not well-defined or documented.

W4. Why is the data labeling process not well-defined or documented?

   A4. Because of a lack of standardization and oversight in data labeling processes.

W5. Why is there a lack of standardization and oversight in data labeling processes?

   A5. Because of a lack of awareness, resources, or investment in data governance.

5. Wrong or no data name convention

W1. Why is there a problem with wrong or no data name convention?

   A1. Because inconsistent, unclear, incorrect, or no data names were entered.

W2. Why inconsistent, unclear, incorrect, or no data names were entered?

A2. Because the person responsible for naming the data forgot or was not aware of the requirement to name the data.

W3. Why did the responsible person forget or not know about the requirement to name the data?

A3. Because of poor communication or training, or because the requirement was not clearly defined or documented.

W4. Why was there poor communication or training, or why was the requirement not clearly defined or documented?

A4. Because of a lack of standardization in data naming processes.

W5. Why is there a lack of standardization in data naming processes?

A5. Because of a lack of awareness, resources, or investment in data governance.

6. Data sheets are not relevant to the company

W1. Why are the data sheets not relevant to the company?

A1. Because the data sheets are either outdated, inaccurate, not aligned with the company's current goals, or not reflective of the company's current processes.

W2. Why are the data sheets outdated, inaccurate, not aligned with the company's current goals, or not reflective of the company's current processes?

A2. Because of a lack of regular maintenance and review of data sheets.

W3. Why is there a lack of regular maintenance and review of data sheets?

A3. Because of a lack of clear guidelines, training, or communication on data maintenance.

W4. Why is there a lack of clear guidelines, training, or communication on data maintenance?

A4. Because of a lack of standardization in data maintenance.

W5. Why is there a lack of standardization in data maintenance?

A5. Because of a lack of awareness, resources or investment in data governance.

7. Multiple data sources

W1. Why are there multiple data sources?

A1. Because different departments use different systems or databases.

W2. Why do different departments use different systems or databases?

A2. Because there has been no consolidation of systems.

W3. Why has there been no consolidation of systems?

A3. Because of a lack of resources, investment, or awareness of the benefit of the consolidation of systems convenience.

W4. Why has there been a lack of resources, investment, or awareness of the benefit of the consolidation of systems?

A4. Because of a lack of centralized data governance.

W5. Why is there a lack of centralized data governance?

A5. Because of a lack of awareness, data-driven culture, leadership and advocacy for centerized data governance initiatives.

8. Difficulties on obtaining wanted data

W1. Why is it difficult to obtain wanted data?

A1. Because the data is not available, not accessible, or the data is not in the desired format or structure.

W2. Why is the data not available, not accessible, or not in the desired format or structure?

A2. Because data storage or data sharing is not consistent for the desired purpose.

W3. Why is data storage or data sharing not consistent for the desired purpose?

A3. Because there is a lack of clear data storage and sharing requirements or guidelines.

W4. Why is there a lack of clear data storage and sharing requirements or guidelines?

A4. Because of a lack of standardization in data storage and sharing.

W5. Why is there a lack of standardization in data storage and sharing?

A5. Because of a lack of awareness, resources, or investment in data governance.

9. Data silos in systems

W1. Why do we have data silos in systems?

A1. Because data is stored in separate, isolated systems.

W2. Why is data stored in separate, isolated systems?

A2. Because there may be technical limitations that prevent different systems from communicating with each other.

W3. Why are there technical limitations?

A3. Because systems or platforms are not compatible with each other.

W4. Why are systems or platforms not compatible with each other?

A4. Because of vendor lock-in, legacy systems, or a lack of universally accepted standards or protocols for interoperability.

W5. Why are there vendor lock-in, legacy systems, or a lack of universally accepted standards or protocols for interoperability?

A5. Because technology providers may intentionally design their systems to be incompatible with others to make it difficult to switch to alternative platforms or technologies. Or because older technologies or systems were developed without considering future compatibility, meanwhile it can be complex and costly to develop and maintain interoperability protocols.

10. Shadow IT

W1. Why do employees use shadow IT?

A1. Because they want to use their own tools or applications.

W2. Why do employees want to use their own tools or applications?

A2. Because the employees want to be more productive and efficient by using their own tools or applications.

W3. Why do employees feel it is more productive and efficient by using their own tools or applications.

A3. Because they may feel that the company's IT systems are too slow, outdated, or restrictive, and they.

W4. Why do employees feel that the company's IT systems are too slow, outdated, or restrictive, and they?

A4. Because the IT department may not have the resources, budget, or expertise to keep up with the latest technology trends and innovations.

W5. Why does the IT department not have the resources, budget, or expertise to keep up with the latest technology trends and innovations?

A5. Because there is a lack of awareness or budget to prioritize IT investments.

11. No requirement of explaining what the purpose of requesting a specific data

W1. Why is there no requirement to explain the purpose of requesting specific data?

A1. Because employees may not understand the importance of providing context for the data request.

W2. Why do employees not understand the importance of providing context for the data request?

A2. Because there is no policy or guideline in place that mandates it.

W3. Why is there no policy or guideline in place that mandates explaining the purpose of requesting specific data?

    A3. Because there is no clear ownership or accountability for data.

W4. Why is there no clear ownership or accountability for data.

    A4. Because the company may not understand the value of data ownership and accountability.

W5. Why does the company not understand the value of data ownership and accountability?

    A5. Because of a lack of leadership or vision around data governance.

12. One end data flow

W1. Why is there a one end data flow?

    A1. Because the company may not have established feedback mechanisms to ensure that data is being used effectively.

W2. Why hasn't the company established feedback mechanisms?

    A2. Because there is no corresponding policy or guideline.

W3. Why is there no corresponding policy or guideline?

        A3. Because there is no clear ownership or accountability for data.

W4. Why is there no clear ownership or accountability for data.

    A4. Because the company may not understand the value of data ownership and accountability.

W5. Why does the company not understand the value of data ownership and accountability?

    A5. Because of a lack of leadership or vision around data governance.

13. Storage data only on personal computer

W1. Why do people store data only on their personal computer?

    A1. Because it is convenient and readily accessible to them.

W2. Why is it convenient to store data only on personal computer?

    A2. Because it allows individuals to have full control over their data and access it at any time without needing an internet connection.

W3. Why do individuals want full control over their data and access it anytime?

   A3. Because there is a lack of clear data storage and sharing requirements or guidelines.

W4. Why is there a lack of clear data storage and sharing requirements or guidelines?

    A4. Because of a lack of standardization in data storage and sharing.

W5. Why is there a lack of standardization in data storage and sharing?

    A5. Because of a lack of awareness, resources, or investment in data governance.

14. Data is not archived or deleted after using

W1. Why is data not archived or deleted after using it?

    A1. Because individuals or organizations may not see the value or importance of doing so.

W2. Why do individuals or organizations not see the value or importance of archiving or deleting data?

    A2. Because they may not have a clear data retention policy or data lifecycle management guidelines in place, or they may lack the knowledge or expertise to implement such policies and guidelines.

W3. Why do individuals or organizations not have a clear data retention policy or data lifecycle management guidelines in place, or lack the knowledge or expertise to implement such policies and guidelines?

    A3. Because they may not have dedicated resources responsible for managing data lifecycle.

W4. Why do individuals or organizations lack the dedicated resources responsible for managing data lifecycle?

    A4. Because they may not prioritize data lifecycle management.

W5. Why do individuals or organizations not prioritize data lifecycle management?

    A5. Because they may not have a clear understanding of the potential risks and consequences of not managing data properly.

15. Nonstandard data sources

W1. Why are there nonstandard data sources?

    A1. Because some data sources may not be properly verified, and the data may not be standardized or normalized.

W2. Why are data sources not properly verified and the data not standardized or normalized?

    A2. Because there may not be a clear understanding of what constitutes reliable data sources or standardization practices.

W3. Why is there a lack of understanding of reliable data sources or standardization practices?

    A3. Because organizations may not have the necessary resources, expertise, or technology to properly verify data sources and standardize data.

W4. Why do organizations lack the necessary resources, expertise, or technology to verify data sources and standardize data?

   A4. Because they may not prioritize standardizing data collection and integration.

W5. Why do organizations not prioritize standardizing data collection and integration?

   A5. Because they may not understand the importance of standardizing data collection and integration or may not have the necessary training or education.

16. Difficulties on using excessive amount of data

W1. Why is it difficult to use excessive amounts of data?

   A1. Because there is too much data to process and analyze.

W2. Why is there too much data to process and analyze?

   A2. Because the data may be unstructured, inconsistent, or contain irrelevant information.

W3. Why is the data unstructured, inconsistent, or contain irrelevant information?

   A3. Because it is gathered from multiple sources, not properly labeled, or lacks clear definitions or categories.

W4. Why is the data gathered from multiple sources, not properly labeled, or lack clear definitions or categories?

   A4. Because there is no standard process for data collection, labeling or classification.

W5. Why is there no standard process for data collection, labeling or classification?

   A5. Because there is no clear ownership or accountability for data within the organization.

17. Time consuming manually compiling work

W1. Why is there time-consuming manually compiling work?

   A1. Some work requires manual compilation and processing of data.

W2. Why does the work require manual compilation and processing of data?

   A2. Because the data is stored in multiple locations, different formats, or is not standardized.

W3. Why is the data stored in multiple locations, different formats, or not standardized?

   A3. Because it is collected and managed by different teams or individuals within the organization, who may have different processes and tools for data processing and analyzing.

W4. Why do different teams or individuals have different processes and tools for data processing and analyzing?

   A4. Because there is no standardization or centralized data governance within the organization.

W5. Why is there no standardization or centralized data governance within the organization?

A5. Because there is no clear ownership or accountability for data, or there is a lack of resources or technology to support data governance. Additionally, there may be limited communication and collaboration between different teams or individuals regarding data governance.

18. IT system old and lack of maintenance

W1. Why is the IT system old?

A1. Because it has not been upgraded for a long time.

W2. Why has the IT system not been updated or upgraded for a long time?

A2. Because of a lack of resources or budget for updating or upgrading the IT system.

W3. Why is there a lack of resources or budget for updating or upgrading the IT system?

A3. Because the organization may not prioritize IT investments.

W4. Why does the organization not prioritize IT investments?

A4. Because of a lack of awareness of the importance of keeping IT systems up to date, or it has not aligned them with business goals.

W5. Why is there a lack of awareness of the importance of keeping IT systems up to date, or it has not aligned them with business goals?

A5. Because of a lack of clear communication channels, collaboration or alignment on priorities, or lack of a strategic roadmap for IT.

19. Information systems are slow

W1. Why are information systems slow?

A1. Because they are taking too long to process requests.

W2. Why are the information systems taking too long to process requests?

A2. Because there is too much traffic on the network.

W3. Why is there too much traffic on the network?

A3. Because the infrastructure is not designed to handle the current volume of requests.

W4. Why is the infrastructure not designed to handle the current volume of requests?

A4. Because the organization may not have anticipated the growth in demand for the system, or because the organization has not invested in upgrading the infrastructure to meet current demands.

W5. Why has the organization not anticipated the growth in demand for the system or invested in upgrading the infrastructure to meet current demands?

A5. Because of a lack of proactive planning and communication, or because of a lack of awareness or prioritization of the importance of investing in IT infrastructure.

20. Data analytics rely on one person's own knowledge and experience

W1. Why does data analytics rely on one person's own knowledge and experience?

   A1. Because that person may have a unique skill set or expertise in a particular area that other team members do not possess.

W2. Why does this unique skill set or expertise result in one person being solely responsible for data analytics?

   A2. Because other team members may not feel comfortable or confident in their ability to effectively analyze and interpret the data.

W3. Why do other team members not feel comfortable or confident in their ability to effectively analyze and interpret the data?

   A3. Because they may not have received sufficient training or have access to the necessary tools and resources to support their analysis.

W4. Why have other team members not received sufficient training or have access to the necessary tools and resources to support their analysis?

   A4. Because of a lack of investment in professional development by the organization.

W5. Why has the organization not invested in professional development to support data analytics?

   A5. Because of a lack of understanding of the potential value and importance of data analytics in decision-making.

21. Inflexible data analytics

W1. Why is the data analytics inflexible?

   A1. Because the current data analytics tools and systems have limited functionalities and do not allow for customization.

W2. Why do the current data analytics tools and systems have limited functionalities?

   A2. Because they were designed to serve a general purpose and cannot be easily customized to fit specific needs.

W3. Why were the data analytics tools and systems designed this way?

   A3. Because it was assumed that a general-purpose tool would be more cost-effective and easier to maintain.

W4. Why would a general-purpose tool be more cost-effective and easier to maintain?

A4. Because customizing tools to fit specific needs requires a lot of updates and maintenance to keep up with changing business needs and technology.

W5. Why can't the data analytics tools and systems be updated and maintained to be more flexible?

A5. Because the cost and effort may be too high.

22. No standardized decision-making principle based on data

W1. Why is there no standardized decision-making principle based on data?

A1. Because there is no clear agreement on what and how data should be used to inform decisions.

W2. Why is there no clear agreement on what and how data should be used to inform decisions?

A2. Because there are different sources of data, and different stakeholders may have different priorities and perspectives on which data is most important, or people do not know how to use data.

W3. Why are there different sources of data and conflicting priorities among stakeholders, or people do not know how to use data?

A3. Because there is no centralized data governance framework that sets standards and guidelines for data collection, management, and usage.

W4. Why is there no centralized data governance framework?

A4. Because of a lack of awareness, resources, or investment in data governance.

W5. Why is there a lack of awareness, resources, or investment in data governance?

A5. Because of a lack of data-driven culture, leadership and advocacy for data governance initiatives, or competing priorities and budget constraints.